%% file: main.tex
\documentclass[conference]{IEEEtran}
\usepackage{cite}

\ifCLASSINFOpdf
  \usepackage[pdftex]{graphicx}
\else
\fi
\usepackage{amsmath}
\usepackage{algorithmic}
\ifCLASSOPTIONcompsoc
  \usepackage[caption=false,font=normalsize,labelfont=sf,textfont=sf]{subfig}
\else
  \usepackage[caption=false,font=footnotesize]{subfig}
\fi
\usepackage{url}

\usepackage{booktabs}
\usepackage{xcolor}
\usepackage{float}
\usepackage{dblfloatfix}
\usepackage{microtype}
\usepackage[shortcuts,acronym]{glossaries}

\usepackage{tikz}
\usepackage{textcomp}
\usepackage{hyperref}
\usepackage{lipsum}

\newcommand\copyrighttext{%
  \footnotesize \textcopyright 2022 IEEE. Personal use of this material is permitted.
  Permission from IEEE must be obtained for all other uses, in any current or future
  media, including reprinting/republishing this material for advertising or promotional
  purposes, creating new collective works, for resale or redistribution to servers or
  lists, or reuse of any copyrighted component of this work in other works.\\
  2022 IEEE International Parallel and Distributed Processing Symposium Workshops (IPDPSW)  978-1-6654-9747-3/22/\$31.00 \textcopyright2022 IEEE\\
  DOI: \href{<http://tex.stackexchange.com>}{10.1109/IPDPSW55747.2022.00024}
  }
  
\newcommand\copyrightnotice{%
\begin{tikzpicture}[remember picture,overlay]
\node[anchor=south,yshift=10pt] at (current page.south) {\fbox{\parbox{\dimexpr\textwidth-\fboxsep-\fboxrule\relax}{\copyrighttext}}};
\end{tikzpicture}%
}

\makeglossaries 
\input{glossaries}

\begin{document}
\bstctlcite{IEEEexample:BSTcontrol}

%
\title{A Hybrid Approach combining ANN-based and Conventional Demapping in Communication for Efficient FPGA-Implementation}

\author{\IEEEauthorblockN{Jonas Ney\IEEEauthorrefmark{2},
Bilal Hammoud\IEEEauthorrefmark{1}, Norbert Wehn\IEEEauthorrefmark{3}}
\IEEEauthorblockA{TU Kaiserslautern\\
Kaiserslautern, Germany\\
Email: \IEEEauthorrefmark{2}ney@eit.uni-kl.de,
\IEEEauthorrefmark{1}hammoud@eit.uni-kl.de,
\IEEEauthorrefmark{3}wehn@eit.uni-kl.de}}



%


\maketitle
\copyrightnotice

\begin{abstract}
In communication systems, \gls{ae} refers to the concept of replacing parts of the transmitter and receiver by \glspl{ann} to train the system end-to-end over a channel model. This approach aims to improve communication performance, especially for varying channel conditions, with the cost of high computational complexity for training and inference. \Glspl{fpga} have been shown to be a suitable platform for energy-efficient \gls{ann} implementation. However, the high number of operations and the large model size of \glspl{ann} limit the performance on the resource-constrained devices, which is critical for low-latency and high-throughput communication systems.
To tackle his challenge, we propose a novel approach for efficient \gls{ann}-based demapping on \glspl{fpga}, which combines the adaptability of the \gls{ae} with the efficiency of conventional demapping algorithms. After adaption to channel conditions, the channel characteristics, implicitly learned by the \gls{ann}, are extracted to enable the use of optimized conventional demapping algorithms for inference. We validate the hardware efficiency of our approach by providing \gls{fpga} implementation results and by comparing the communication performance to that of conventional systems.
Our work opens a door for the practical application of \gls{ann}-based communication algorithms on \glspl{fpga}.  

\end{abstract}
\glsresetall


%
\IEEEpeerreviewmaketitle

\input{sections/01_introduction}

\input{sections/02_methodology}

\input{sections/04_results}

\input{sections/05_conclusion}
\section*{Funding}
This work was partly funded by the German ministry of education and research (BMBF) under grant 16KIS1185 (FunKI) and under grant 16KISK004 (Open6GHuB).


\bibliographystyle{IEEEtran}
\bibliography{bibliography}

\end{document}

%% file: glossaries.tex
\newcommand{\newacronymf}[3]{\newglossaryentry{#1}{name={#2},description={#3},first={#2}}}

\newacronymf{4g}{4G}{fourth generation of mobile phone system specifications}
\newacronymf{5g}{5G}{fifth generation of mobile phone system specifications}
\newacronym{1dlstm}{1D-LSTM}{one-dimensional long short-term memory}
\newacronym{2dlstm}{2D-LSTM}{2D long Short-term memory}
\newacronym{1dcnn}{1D-CNN}{1D convolutional neural network}
\newacronym{2dcnn}{2D-CNN}{2D convolutional neural network}

\newacronym{ann}{ANN}{artificial neural network}
\newacronym{app}{APP}{a posteriori probability}
\newacronym{ask}{ASK}{amplitude-shift keying}
\newacronym{awgn}{AWGN}{additive white Gaussian noise}
\newacronym{arq}{ARQ}{automatic repeat request}
\newacronym{acs}{ACS}{add-compare-select}
\newacronym{asic}{ASIC}{application specific integrated circuit}
\newacronym{arp}{ARP}{almost regular permuatation}
\newacronym{asip}{ASIP}{application specific instruction set processor}
\newacronymf{amba}{AMBA}{advanced microcontroller bus architecture}
\newacronym{ahb}{AHB}{advanced high-performance bus}
\newacronym{aacs}{AACS}{a priori aware constant sphere}
\newacronym{ant}{ANT}{algorithmic noise-tolerance}
\newacronym{apu}{APU}{application processing unit}
\newacronym{ae}{AE}{Autoencoder}

\newacronym{blstm}{BLSTM}{bidirectional long short-term memory}
\newacronym{bpsk}{BPSK}{binary phase shift keying}
\newacronym{ber}{BER}{bit error rate}
\newacronym{bp}{BP}{belief propagation}
\newacronym{bw}{BW}{bit width}
\newacronym{bicm}{BICM}{bit interleaved coded modulation}
\newacronym{bler}{BLER}{block error rate}
\newacronym{bilstm}{Bi-LSTM}{bidirectional LSTM}
\newacronym{bimc}{BIMC}{bit interleaved coded modulation}
\newacronym{bmd}{BMD}{bit-metric decoding}
\newacronym{bram}{BRAM}{block RAM}

\newacronym{ctc}{CTC}{connectionist temporal classification}
\newacronym{cn}{CN}{check node}
\newacronym{cr}{CR}{code rate}
\newacronym{cfu}{CFU}{Check Node Functional Unit}
\newacronym{crc}{CRC}{cyclic redundancy check}
\newacronym{cc}{CC}{convolutional code}
\newacronym{cnb}{CNB}{check node block}
\newacronym{cnp}{CNP}{check node processor}
\newacronym{cdma}{CDMA}{code division multiple access}
\newacronymf{cmos}{CMOS}{complementary metal oxide semiconductor}
\newacronymf{cnn}{CNN}{convolutional neural network}
\newacronymf{cer}{CER}{character error rate}
\newacronym{cpu}{CPU}{central processing unit}
\newacronym{cgp}{CGP}{cartesian genetic programming}

\newacronym{darts}{DARTS}{differentiable architecture search}
\newacronym{dnnu}{DNNU}{deep neural network unit}
\newacronym{dma}{DMA}{direct memory access}
\newacronym{db}{dB}{decibel}
\newacronym{dp}{DP}{dual ported (memory)}
\newacronym{dram}{DRAM}{dynamic random access memory}
\newacronym{dc}{$d_c$}{CN degree}
\newacronym{dv}{$d_v$}{VN degree}
\newacronym{dvs}{DVS}{dynamic voltage scaling}
\newacronym{dvb}{DVB}{digital video broadcast}
\newacronymf{dvbrcs}{DVB-RCS}{digital video broadcasting - return channel over satellite}
\newacronymf{dvbrct}{DVB-RCT}{digital video broadcasting - return channel terrestrial}
\newacronymf{dvbc2}{DVB-C2}{digital video broadcasting - cable - second generation \cite{eudigital}}
\newacronymf{dvbs2}{DVB-S2}{digital video broadcasting - satellite - second generation \cite{eudigital}}
\newacronymf{dvbs2x}{DVB-S2X}{digital video broadcasting - satellite - extension of the second generation \cite{eudigital}}
\newacronymf{dvbt2}{DVB-T2}{digital video broadcasting - terrestrial - second generation \cite{eudigital}}
\newacronym{dvfs}{DVFS}{dynamic voltage and frequency scaling}
\newacronym{drp}{DRP}{dithered relative prime}
\newacronym{dnn}{DNN}{deep neural network}
\newacronym{dl}{DL}{deep learning}
\newacronym{dsp}{DSP}{digital signal processor}
\newacronym{dspu}{DSP-unit}{digital signal processor unit}
\newacronym{dop}{DOP}{degree of parallelism}
\newacronym{dr}{DR}{decision region}

\newacronym{esf}{ESF}{extrinsic scaling factor}
\newacronym{edc}{EDC}{error detection and correction}
\newacronym{edge}{EDGE}{name of a communications standard}
\newacronym{eds}{EDS}{error detection sequential}
\newacronym{ecc}{ECC}{error correction code}
\newacronym{exit}{EXIT}{extrinsic information transfer}

\newacronym{fsm}{FSM}{finite-state machine}
\newacronym{fec}{FEC}{forward error correction}
\newacronym{fer}{FER}{frame error rate}
\newacronymf{fifo}{FIFO}{first in first out memory}
\newacronym{fcn}{FCN}{fully convolutional network}
\newacronym{fpn}{FPN}{feature pyramid network}
\newacronym{fpga}{FPGA}{field-programmable gate array}

\newacronym{gan}{GAN}{generative adversarial network}
\newacronym{gpp}{GPP}{general-purpose processor}
\newacronym{gprs}{GPRS}{name of a communications standard}
\newacronym{gps}{GPS}{global positioning system}
\newacronymf{geq}{GE}{gate equivalent}
\newacronym{gfq}{GF$(q)$}{galois field}
\newacronym{gru}{GRU}{gated recurrent unit}
\newacronym{gpu}{GPU}{graphics processor unit}
\newacronym{gops}{GOPS}{giga operations per second}

\newacronymf{hspa}{HSPA}{high speed packet access}
\newacronym{half}{HALF}{\underline{H}olistic \underline{A}uto machine \underline{L}earning for \underline{F}PGAs}
\newacronym{harq}{HARQ}{hybrid automatic repeat request}
\newacronym{hw}{HW}{hardware}
\newacronym{hls}{HLS}{high-level synthesis}
\newacronym{hpc}{HPC}{high performance computing}

\newacronym{ic}{IC}{integrated circuit}
\newacronym{idd}{IDD}{iterative demapping and decoding}
\newacronym{io}{I/O}{input / output}
\newacronym{ip}{IP}{intellectual property}
\newacronym{ier}{IER}{injected error rate}
\newacronym{imdd}{IM/DD}{intensity modulation/direct detection}
\newacronym{iot}{IoT}{internet of things}
\newacronym{itu}{ITU}{international telecommunication union}

\newacronym{knn}{KNNs}{künstliche neuronale Netze}

\newacronym{lstm}{LSTM}{long short-term memory}
\newacronym{llr}{LLR}{logarithmic likelihood ratio}
\newacronym{ldr}{LDR}{Logarithmic Density Ratio}
\newacronymf{lte}{LTE}{long term evolution of the 3GPP standard}
\newacronym{ldpc}{LDPC}{low-density parity-check}
\newacronymf{logmap}{Log-MAP}{MAP algorithm in the logarithmic domain}
\newacronym{lan}{LAN}{local area network }
\newacronym{lsb}{LSB}{least significant bit}
\newacronym{lut}{LUT}{look-up table}
\newacronym{lfsr}{LFSR}{linear feedback shift register}
\newacronym{lord}{LORD}{layered orthogonal lattice detector}
\newacronym{lsd}{LSD}{List sphere decoder}
\newacronym{ldmc}{LDMC}{low-density MIMO check}

\newacronym{mac}{MAC}{multiply-accumulate}
\newacronym{map}{MAP}{maximum a posteriori probability}
\newacronymf{mlm}{Max-Log-MAP}{MAP algorithm in the logarithmic domain with simplified computation}
\newacronym{mimo}{MIMO}{multiple input multiple output}
\newacronym{mtbf}{MTBF}{mean time between failure}
\newacronym{msb}{MSB}{most significant bit}
\newacronym{mops}{MOPS}{million operations per second}
\newacronym{mmse}{MMSE}{minimum mean square error}
\newacronym{mlh}{ML}{maximum likelihood}
\newacronym{ml}{ML}{machine learning}
\newacronym{mdlstm}{MD-LSTM}{multidimensional long short-term memory}
\newacronym{mlp}{MLP}{multilayer perceptron}
\newacronym{mi}{MI}{mutual information}
\newacronym{mpsoc}{MPSOC}{multiprocessor system on a chip}

\newacronym{nas}{NAS}{neural architecture search}
\newacronym{nns}{NNs}{neural networks}
\newacronym{nn}{NN}{neural network}
\newacronym{nbldpc}{NB-LDPC}{non-binary low-density parity-check}
\newacronym{nii}{NII}{next iteration initialization}
\newacronym{nsc}{NSC}{non-systematic and non-recursive convolutional}
\newacronym{noc}{NoC}{network on chip}
\newacronymf{nom}{NOM}{nominal case, 1.2V, 25$^\circ$C}

\newacronym{ocr}{OCR}{optical character recognition}
\newacronym{otx}{OTX}{outer receiver}
\newacronym{ofdm}{OFDM}{orthogonal frequency division multiplexing}

\newacronym{pl}{PL}{programmable logic}
\newacronym{ps}{PS}{processing system}
\newacronym{psk}{PSK}{phase-shift keying}
\newacronym{psmap}{PSMAP}{parallel SMAP}
\newacronym{pr}{P\&R}{placement and routing}
\newacronym{ped}{PED}{partial euclidean distance}
\newacronym{pic}{PIC}{parallel interference cancellation}
\newacronym{pn}{PN}{permutation node}
\newacronym{pe}{PE}{processing element}

\newacronym{qam}{QAM}{quadrature amplitude modulation}
\newacronym{qpsk}{QPSK}{quadrature phase shift keying}
\newacronym{qpp}{QPP}{quadratic permutation polynomial}
\newacronym{qos}{QoS}{quality of service}

\newacronym{relu}{ReLU}{rectified linear unit}
\newacronym{rl}{RL}{reinforcement learning}
\newacronym{rnns}{RNNs}{recurrent neural networks}
\newacronym{rnn}{RNN}{recurrent neural network}
\newacronym{rsc}{RSC}{recursive systematic convolutional}
\newglossaryentry{ram}{name=RAM,description=random access memory,first=random access memory (RAM),firstplural=random access memories (RAMs)}
\newglossaryentry{rom}{name=ROM,description=read only memory,first=read only memory (ROM),firstplural=read only memories (ROMs)}
\newacronym{rms}{RMS}{recognition, mining, and synthesis}
\newacronym{rs}{RS}{reed-solomon code}
\newacronym{rtl}{RTL}{register transfer level}
\newacronym{rts}{RTS}{repeated tree search}
\newacronym{rap}{RAP}{resilience articulation point}

\newacronym{snr}{SNR}{signal-to-noise ratio}
\newacronym{smap}{SMAP}{serial MAP}
\newacronym{sdr}{SDR}{software-defined radio}
\newacronymf{sdram}{SDRAM}{synchronous dynamic random access memory}
\newacronym{set}{SET}{single-event transient}
\newacronym{seu}{SEU}{single-event upset}
\newacronym{siso}{SISO}{soft-input soft-output}
\newacronym{siso2}{SISO}{single-input single-output}
\newacronym{soc}{SoC}{systems-on-chip}
\newacronym{sova}{SOVA}{soft-output Viterbi algorithm}
\newacronym{sw}{SW}{software}
\newacronym{sm}{SM}{spatial multiplexing}
\newacronym{stbc}{STBC}{space-time block code}
\newacronym{sgd}{SGD}{stochastic gradient descend}
\newacronym{sttc}{STTC}{space-time trellis code}
\newacronym{sic}{SIC}{successive interference cancellation}
\newacronym{sts}{STS}{single tree search}
\newacronym{synb}{SYN}{syndrome-based}
\newacronym{simd}{SIMD}{single instruction multiple data}
\newacronym{sota}{SOTA}{state of the art}

\newacronym{tdp}{TDP}{thermal design power}
\newacronym{tmr}{TMR}{triple modular redundancy}
\newacronym{tc}{TC}{turbo code}
\newacronym{tpu}{TPU}{tensor processing unit}

\newacronymf{umts}{UMTS}{universal mobile telecommunications system}
\newacronymf{uwb}{UWB}{ultra-wide band}
\newacronym{usb}{USB}{universal serial bus}
\newacronym{umic}{UMIC}{ultra high speed mobile information and communication}
\newacronym{urllc}{uRLLC}{ultra reliable low latency communication}

\newacronym{vn}{VN}{variable node}
\newacronym{vfu}{VFU}{variable node functional unit}
\newacronym{vhdl}{VHDL}{very high speed integrated circuits hardware description language}
\newacronym{vliw}{VLIW}{very large instruction word}
\newacronym{vnb}{VNB}{variable node block (in \gls{LDPC} decoder architectures)}
\newacronym{vnp}{VNP}{variable node processors (in \gls{LDPC} decoder architectures)}

\newacronymf{wer}{WER}{word error rate}

\newacronym{xor}{XOR}{exclusive OR function}

\newacronym{zf}{ZF}{zero-forcing}

%% file: sections/01_introduction.tex
\section{Introduction}
\label{sec:introduction}

In order to satisfy the increasing communication demand of embedded devices and facilitate further advancements in fields like autonomous driving, efficient hardware implementation of communication systems is of great importance. In the era of 5G and 6G, three major challenges are associated with the design of communication algorithms on embedded devices: \textbf{low-latency} is essential for delay-sensitive communication, \textbf{low-power} is mandatory to meet the energy constraints of embedded applications, and near real-time \textbf{adaptability} is of high relevance due to \textit{varying channel conditions} between communication entities.

In the context of communication, the \gls{ae} approach tackles the challenge of designing an end-to-end system that can adapt to channel fluctuations in near real-time. It was first proposed in \cite{shea:2017}, where parts of the classical transmitter and receiver are replaced by \glspl{ann} to train all components jointly, optimizing the performance of the global system. The approach of an \gls{ae} for mapping and demapping of communication symbols was first validated for a real-world channel in \cite{doerner2018} via over-the-air measurements using \glspl{sdr}. To resolve imperfections of the channel model and especially to adapt to varying channel conditions, retraining of the demapper-\gls{ann} was performed over the real channel after initial end-to-end training for an abstract channel model like \gls{awgn}.


Although \glspl{ann} are able to improve the performance of traditional communication systems \cite{EndToEndDeepLearningForOpticalFiberCommunications, TrainableCommunicationSystems}, they are challenging in terms of efficient hardware implementation. Whenever dealing with \gls{ann}-based methods, their high memory demand and computational requirements yield to large latency, power and energy consumption of the hardware implementation. These challenges are very significant for resource-constrained communication systems. In addition, conventional algorithms are highly optimized for efficient implementation. Because of those reasons, \gls{ann} solutions had no fundamental impact on the practical implementation of baseband signal processing so far.
In this paper we focus on demapping, which is an important task in baseband signal processing and largely impacts the overall communication performance. 

\Glspl{fpga} have been shown to be well suited for efficient implementation of \glspl{ann} in terms of power, latency, and reconfigurability \cite{gaikwad2019, Venieris2017}. However, the majority of \gls{fpga}-based \gls{ann} architectures target the inference only, by training the network in software and exporting the trained parameters to optimized inference engines  \cite{umuroglu2017, zhang2018}. This method is not sufficient for the application of \glspl{ae} in our context, as the main advantage of the trainable system is the adaptability to varying channel conditions in real-time. Therefore, an \gls{ann} \gls{fpga} implementation, which enables real-time training becomes mandatory for our application.

Due to the reconfigurability of \glspl{fpga}, it is possible to fully exploit the available resources for inference until retraining is performed, and then utilize all resources for efficient training. 
While the \gls{ann} training algorithm is of higher complexity than the inference, the inference is performed much more frequently. Thus inference optimization is of high relevance to reduce power, energy and latency of the system. 

Therefore, the novelties of this paper are
\begin{itemize}
    \item A new hybrid approach that combines the adaptability of \glspl{ann} with conventional algorithms, by extracting the learned channel characteristics of the trained-demapper. In this way we combine the efficiency of low complex conventional demappers with the advantages of adjusting to changing channel condition. 
    \item We present detailed implementation results of the new approach on a Xilinx ZCU3EG \gls{fpga} and compare it to conventional demapping algorithms. 
\end{itemize}

%% file: sections/02_methodology.tex
\section{Hybrid approach}

\begin{figure}[!h]
\centering
\includegraphics[width=\columnwidth]{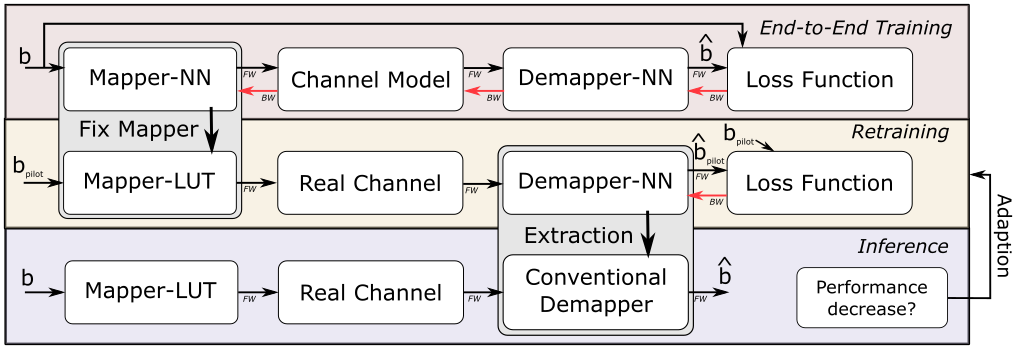}
\caption{Hybrid approach}
\label{fig:methodology}
\end{figure}

Our novel approach for the efficient implementation of \gls{ae}-based demappers allows to utilize the benefits of the \gls{ae} concept for adapting to varying channel conditions and simultaneously enables the use of highly sophisticated conventional demapping algorithms for inference as shown in Fig. \ref{fig:methodology}. 

Our approach can be divided into multiple steps:
\begin{enumerate}
    \item \textbf{E2E Training:} Training of the end-to-end system including transmitter and receiver \glspl{ann} for an abstract channel model in software.
    \item \label{item:retraining} \textbf{Retraining:} Retrain the system for real channel conditions by fixing the mapper constellation and retraining the demapper \gls{ann} in hardware to compensate for imperfections of the channel model.
    \item \label{item:inference} \textbf{Inference:} Enable the use of conventional demapping algorithms for efficient inference by extracting the decision regions from the demapper \gls{ann}. If communication performance suffers due to varying channel conditions, trigger retraining.
\end{enumerate}

The steps of our approach are described in more detail in the following. 

\subsection{E2E Training}
In traditional communication systems, transmitter and receiver comprise several building blocks (e.g. encoder, decoder, mapper, demapper) to overcome the signal distortion and noise introduced by the channel. In the \textit{E2E Training} step, we focus on replacing the traditional mapping and demapping blocks by \glspl{ann} to learn a complex symbol representation that is robust with respect to the noise of an abstract channel model. The mapper defines a building block that takes an input vector consisting of $m$ bits and maps it to a complex-valued constellation symbol, which is then transmitted over a channel. The goal of the mapper is to find a symbol constellation that conveys the maximum amount of information at the given system settings. Subsequently, the demapper receives the corresponding complex symbol that is impaired by the noise of the channel. Its goal is to provide the most reliable estimate of the originally transmitted bits in form of $m$ output probabilities. 
For our approach, the \gls{ann} topologies of mapper and demapper are constrained to layers without memory (e.g. \glspl{rnn}) as the input space of the demapper needs to be two-dimensional. Apart from that, the structure of the networks can be chosen arbitrarily.
In the \textit{E2E Training} step, those networks are trained for a specific number of epochs to increase the bitwise \gls{mi} by minimizing the binary cross-entropy loss. Using this approach the \glspl{ann} are able to learn non-uniform constellations which increase the bitwise \gls{mi} as compared to conventional \gls{qam} constellations for the underlying channel model as shown in \cite{TrainableCommunicationSystems}.

\subsection{Retraining}

As the abstract channel model used for end-to-end training is not able to perfectly emulate all properties of a real channel which might also vary over time, it is necessary to adapt to changing channel conditions during runtime. Therefore, we propose an \gls{fpga} implementation of a trainable \gls{ann} for the demapper of the communication system. To reduce the bandwidth and eliminate the communication overhead of a feedback channel, we fix the constellations of the transmitter \gls{ann} after the \textit{E2E Training}. This way the receiver can adapt to fluctuations of the channel as shown in \cite{doerner2018} where finetuning of the receiver improved the performance of over-the-air transmission by up to 1 dB.
Thus, our approach is based on a low-power \gls{fpga} implementation of a trainable \gls{ann} for demapping that can be retrained for changing channel conditions. 
We implement the forward and the backward path of the \gls{ann} as a pipelined architecture using Vivado \gls{hls} and Vivado Design Suite 2019.2. To reduce latency and increase throughput, separate hardware modules for each \gls{ann}-layer are instantiated. The layers are based on Xilinx FINN library \cite{umuroglu2017} and allow for flexible adjustment of the \gls{dop} which enables to trade-off between latency and power consumption.

\subsection{Inference}

After retraining is performed and the network has adapted to the channel, we extract the learned demapping characteristics from the \gls{ann} as follows: first, we sample over the two-dimensional input space of the demapper-\gls{ann} to get the learned symbol (\gls{ann}-output) for each complex input sample (\gls{ann}-input). This gives us the \glspl{dr} of each symbol, as described later in Fig. \ref{fig:dr_illustragion}. Since this \gls{dr}-diagram can be interpreted as Voronoi diagram, we can find a centroid $c_i$ for each Voronoi cell corresponding to the symbol $s_i$ such that $c_i$ represents all points $X$ whose distance to $c_i$ is less than their distance to all other centroids. The centroids are calculated based on the vertices of each Voronoi cell and can be used by conventional demapping algorithms for efficient inference. It is to note that those centroids do not necessarily replicate the constellation of the mapper but implicitly include the learned information of the \gls{ann} to compensate channel impairments, e.g. in Fig. \ref{fig:dr_illustragion} they include the phase-shift of the channel.  

To detect changes in channel conditions, the performance of the system can be regularly evaluated, either by periodically sending pilot symbols to trigger retraining of the demapper if the \gls{ber} reaches a threshold or by using an outer \gls{ecc} as demonstrated in \cite{schibisch2018}. The number of bit flips that are corrected by the \gls{ecc} can guide as performance metric of the communication system and therefore detect environmental changes and activate retraining.

%% file: sections/04_results.tex
\section{Results}

\subsection{Setup}

To evaluate the efficiency of our approach we present a case study for a communication system based on 16-\gls{qam} modulation. We jointly train the mapper and demapper \gls{ann} over an \gls{awgn} channel in PyTorch for different \glspl{snr} to obtain an optimized constellation for each \gls{snr}. The mapper consists of a trainable embedding layer with 16 inputs and two outputs as well as an average power normalization layer to map each input symbol to a complex constellation point. The demapper \gls{ann} has two inputs and is composed of three fully connected layers with 16 neurons each, followed by a \gls{relu} layer and a final sigmoid layer to receive output probabilities for each of the four bits. 
After initial end-to-end training in Python, the demapper is implemented as trainable \gls{ann} on the Avnet Ultra96-V2 featuring the Xilinx ZU3EG \gls{fpga}. Furthermore, the algorithm to extract the centroids from the trained demapper, as well as a conventional soft-demapping algorithm, are implemented on the board. To reduce the complexity of the demapping, we implement the suboptimal soft-demapping algorithm proposed in \cite{softDemapping} which replaces exponential and logarithmic functions by simplified calculations according to:  
\begin{equation*}
	\small
	llr(b_k|s_r) =  \frac{1}{2 \sigma^2}. \left\{ \min_{i}(s_r-c_{i,k=0})^2 -  \min_i(s_r-c_{i,k=1})^2\right\}  
\end{equation*}
where $b_k$ is the $k$-th bit, $s_r$ is the received symbol and $c_{i,k=0}$ and $c_{i,k=1}$ represent the centroids of region with label $i \in \{1:M\}$ whose $k$-th bits are `0' and `1' respectively.

To verify the adaptability of the system and to illustrate changing environmental conditions we select a channel that introduces a fixed phase-offset with respect to the \gls{awgn} channel of the end-to-end training.   

\subsection{Communication Performance Comparison}

\begin{figure}[!h]
	\centering
	\includegraphics[width=0.8\columnwidth]{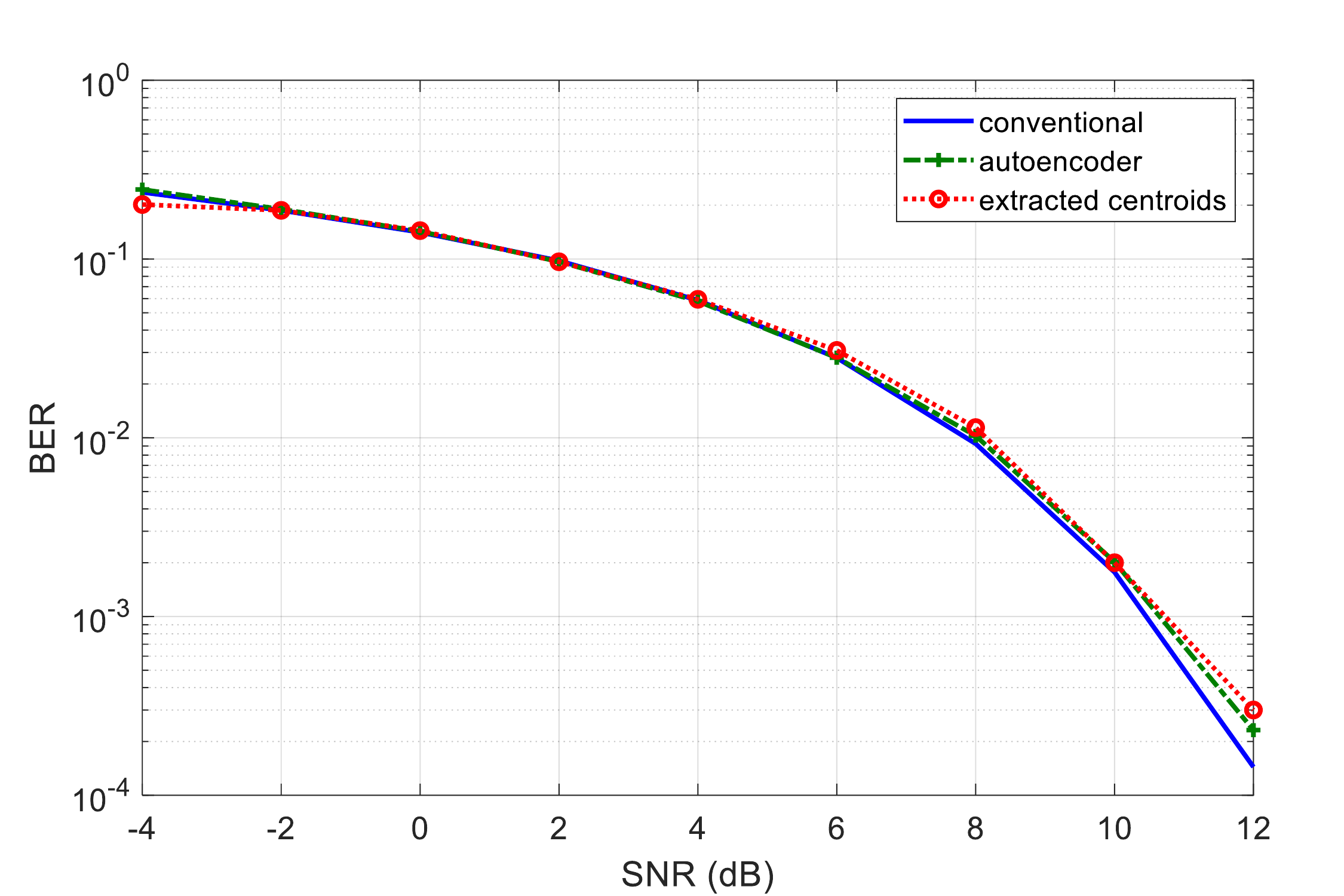}
	\caption{Bit-error-rate (BER) of different demapping algorithms.}
	\label{fig:ber_snr}
\end{figure}

To evaluate the communication performance of our approach, we train the \gls{ae} in end-to-end manner over an \gls{awgn} channel for different \glspl{snr}, resulting in different constellations for each \gls{snr}. Afterward, the learned parameters of the demapper-\gls{ann} are mapped to our hardware architecture for the different \glspl{snr}. Subsequently, we apply our algorithm to extract the learned centroids from the trained-\glspl{ann} and apply suboptimal soft-demapping based on those centroids. The \gls{ber} of this approach is compared to that of a conventional soft-demapping algorithm and to the \gls{ber} using the \gls{ae} for inference, as shown in Fig. \ref{fig:ber_snr}.
It can be seen that the performance of the \gls{ae} and the extracted centroids is on the level of the conventional demapper for \glspl{snr} up to 10 dB and only decreases slightly for 12 dB. Those results validate the feasibility of our extraction algorithm with respect to communication performance. The corresponding hardware efficiency of our approach is investigated in Sec. \ref{sec:HardwarePerformanceEvalutaion}.

\subsection{Adaptability Illustration}
\begin{figure}[H]
\centering
\includegraphics[width=1\columnwidth]{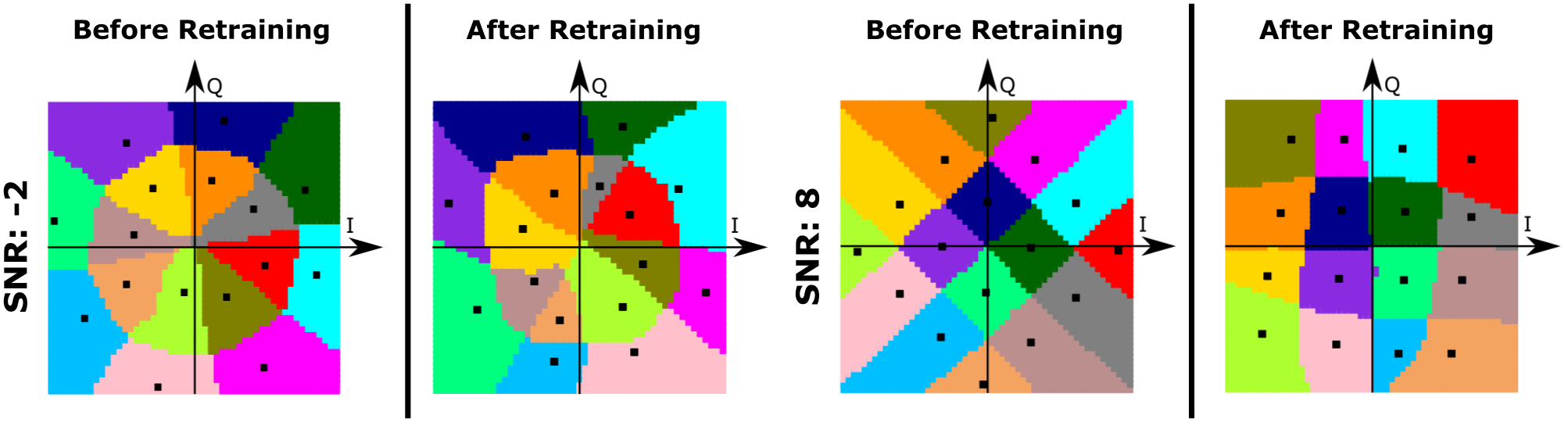}
\caption{\gls{dr} and centroid illustration}
\label{fig:dr_illustragion}
\end{figure}

To illustrate the adaptability of our approach to channel fluctuations, we use a \gls{awgn} channel with 0 phase-shift for end-to-end training. Then, the \glspl{ann} are mapped to \gls{fpga} and the demapper-\gls{ann} is retrained for a channel with a phase-shift of $\frac{\pi}{4}$ on the board. Afterward, the centroids of the retrained demapper are extracted and applied to the suboptimal soft-demapping algorithm. The \gls{dr} of the demapper and the corresponding centroids are shown in Fig. \ref{fig:dr_illustragion} before and after retraining for an \gls{snr} of -2 and 8 dB respectively. 
It can be seen that for both \glspl{snr} the \glspl{dr} are rotated by $\frac{\pi}{4}$ after retraining. 
This results in an improved communication performance for the \gls{ae}-inference and the conventional algorithm applied to the extracted centroids as given in Tab. \ref{tab:phase_offset_adaption}. 

\begin{table*}[!ht]
\renewcommand\thetable{2}
\centering
\caption{Comparison of \gls{ae}-based inference to conventional soft demapping}
\label{tab:comparison_ae_conventional}
\begin{tabular}{ccccccccc}
\toprule

&
Latency [s] &
Throughput [symbols/s] &
BRAM &
DSP &
FF &
LUT &
Power [W] &
Energy [J/symbol] \\

\midrule


\begin{tabular}[c]{@{}c@{}}Soft-demapper with learned centroids\end{tabular} &
5.33$\mathbf{\cdot 10^{-8}}$    &
7.50$\mathbf{\cdot 10^{7}}$     &
0                               &
1                               &
1042                            &
1107                            &
5.5$\mathbf{\cdot 10^{-2}}$     &
7.33$\mathbf{\cdot 10^{-10}}$   \\

\gls{ae}-inference              &
8.10$\mathbf{\cdot 10^{-8}}$    &
1.23$\mathbf{\cdot 10^{7}}$     &
18.5                            &
352                             &
10895                           &
11343                           &
4.53$\mathbf{\cdot 10^{-1}}$    &
3.67$\mathbf{\cdot 10^{-8}}$    \\

\midrule 

\gls{ae}-training               &
2.67$\mathbf{\cdot 10^{-7}}$    &
3.75$\mathbf{\cdot 10^{6}}$     &
89                              &
343                             &
19013                           &
19793                           &
5.47$\mathbf{\cdot 10^{-1}}$    &
1.46$\mathbf{\cdot 10^{-7}}$    \\

\bottomrule
\end{tabular}
\end{table*}

\begin{table}[H]
\renewcommand\thetable{1}
\centering
\caption{Phase-Offset adaption of \gls{ae} and conventional algorithm applied to extracted centroids}
\label{tab:phase_offset_adaption}
\begin{tabular}{cccccc}
\toprule

&
&
\multicolumn{2}{c}{Before Retraining} &
\multicolumn{2}{c}{After Retraining} \\

\gls{snr}                       &
\begin{tabular}[c]{@{}c@{}}Baseline\\\gls{ber}\end{tabular} &
\begin{tabular}[c]{@{}c@{}}\gls{ae}\\\gls{ber}\end{tabular} &
\begin{tabular}[c]{@{}c@{}}Cent. extraction\\\gls{ber}\end{tabular} &
\begin{tabular}[c]{@{}c@{}}\gls{ae}\\\gls{ber}\end{tabular} &
\begin{tabular}[c]{@{}c@{}}Cent. extraction\\\gls{ber}\end{tabular} \\

\midrule 

-2      &
0.19    &
0.318   &
0.319   &
0.199   &
0.2005  \\

8      &
0.0103    &
0.316   &
0.323   &
0.0127   &
0.0143  \\

\bottomrule
\end{tabular}
\end{table}

The baseline \gls{ber} gives the lower bound, corresponding to a channel without any phase-offset. The upper bound is given by the \glspl{ber} before retraining, corresponding to a conventional algorithm without any adaption to channel fluctuations. It can be observed that the \glspl{ber} after retraining nearly approach the baseline \gls{ber} and thus the phase-shift is nearly fully compensated. This is the case for the \gls{ae}-inference and the demapping based on extracted centroids which implies that there is no drawback of using the extracted centroids as compared to the \gls{ae}-inference with respect to communication performance for the evaluated channel fluctuation.

\subsection{\gls{fpga} Results}
\label{sec:HardwarePerformanceEvalutaion}

The inference of the demapper-\gls{ann} is designed to achieve maximal resource utilization on the Xilinx ZCU3EG, limited by the amount of available DSPs. For a fair comparison, a soft-demapping algorithm is implemented with latency and throughput in the same order of magnitude as the \gls{ann}-inference. Furthermore, both designs achieve the same communication performance, as demonstrated in the previous sections. Tab. \ref{tab:comparison_ae_conventional} shows a comparison for both implementations in terms of hardware performance (latency, throughput), resource utilization and efficiency (power, energy). The soft-demapper highly outperforms the \gls{ae} regarding resource utilization, power consumption and energy efficiency. It has 10$\times$ lower LUT-usage and utilizes 352$\times$ less DSP, resulting in 10$\times$ reduced power consumption and  50$\times$ higher energy efficiency. This allows for performing demapping in parallel by instantiating multiple modules of the soft-demapper to approach a throughput in the order of \textit{Gbps}, which could not be accomplished with the \gls{ae}-inference, due to the limited amount of resources.
Those results support our novel approach, since we maintain the communication performance of the \gls{ae} while being much more hardware efficient, resulting in either reduced resource consumption or higher throughput for fixed amount of resources.

In Tab. \ref{tab:comparison_ae_conventional}, we also show \gls{fpga} implementation results of the \gls{ann} training module. It can be seen that the resource utilization is much higher compared to the inference, due to the increased complexity. As the inference is performed much more frequently , this would result high idle time of the training module on an \gls{asic}. In contrast, \gls{fpga} can be reconfigured to either perform training or inference, resulting in a more efficient use of resources. This is why \glspl{fpga} are very well suited for this application and our hybrid approach.

%% file: sections/05_conclusion.tex
\section{Conclusion}

We proposed a hybrid approach for the efficient implementation of \gls{ae}-based demapping algorithms on \gls{fpga}. The novel approach is based on a trainable-\gls{ann} \gls{fpga} architecture which is utilized to retrain for channel fluctuations during runtime. Afterward, the learned information is extracted in form of Voronoi-centroids, which allows to apply optimized conventional demapping algorithms for inference. The approach enables the use of the adaptable \gls{ae} concept while preserving the efficiency of conventional algorithms.  
We evaluated the communication performance and hardware characteristics of our approach by presenting corresponding hardware implementation results. Therefore we applied suboptimal soft-demapping to our extracted centroids and validated that a phase-offset of $\frac{\pi}{4}$ can be compensated by retraining of the \gls{ann}. This showed that our hybrid approach combines the advantages of \glspl{ae} in terms of communication performance with the efficiency of highly optimized conventional demapping algorithms.   

%% file: main.bbl
\begin{thebibliography}{10}
\providecommand{\url}[1]{#1}
\csname url@samestyle\endcsname
\providecommand{\newblock}{\relax}
\providecommand{\bibinfo}[2]{#2}
\providecommand{\BIBentrySTDinterwordspacing}{\spaceskip=0pt\relax}
\providecommand{\BIBentryALTinterwordstretchfactor}{4}
\providecommand{\BIBentryALTinterwordspacing}{\spaceskip=\fontdimen2\font plus
\BIBentryALTinterwordstretchfactor\fontdimen3\font minus
  \fontdimen4\font\relax}
\providecommand{\BIBforeignlanguage}[2]{{%
\expandafter\ifx\csname l@#1\endcsname\relax
\typeout{** WARNING: IEEEtran.bst: No hyphenation pattern has been}%
\typeout{** loaded for the language `#1'. Using the pattern for}%
\typeout{** the default language instead.}%
\else
\language=\csname l@#1\endcsname
\fi
#2}}
\providecommand{\BIBdecl}{\relax}
\BIBdecl

\bibitem{shea:2017}
T.~O’Shea and J.~Hoydis, ``An introduction to deep learning for the physical
  layer,'' \emph{IEEE Transactions on Cognitive Communications and Networking},
  vol.~3, no.~4, pp. 563--575, 2017.

\bibitem{doerner2018}
S.~D{\"{o}}rner \emph{et~al.}, ``Deep learning based communication over the
  air,'' \emph{IEEE Journal of Selected Topics in Signal Processing}, vol.~12,
  no.~1, pp. 132--143, 2018.

\bibitem{EndToEndDeepLearningForOpticalFiberCommunications}
\BIBentryALTinterwordspacing
B.~Karanov \emph{et~al.}, ``End-to-end deep learning of optical fiber
  communications,'' \emph{Journal of Lightwave Technology}, vol.~36, no.~20, p.
  4843–4855, Oct 2018. [Online]. Available:
  \url{http://dx.doi.org/10.1109/JLT.2018.2865109}
\BIBentrySTDinterwordspacing

\bibitem{TrainableCommunicationSystems}
S.~Cammerer \emph{et~al.}, ``Trainable communication systems: Concepts and
  prototype,'' \emph{IEEE Transactions on Communications}, vol.~68, no.~9, pp.
  5489--5503, 2020.

\bibitem{gaikwad2019}
N.~B. Gaikwad \emph{et~al.}, ``Efficient {FPGA} implementation of multilayer
  perceptron for real-time human activity classification,'' \emph{IEEE Access},
  vol.~7, pp. 26\,696--26\,706, 2019.

\bibitem{Venieris2017}
S.~I. Venieris and C.-S. Bouganis, ``Latency-driven design for fpga-based
  convolutional neural networks,'' in \emph{2017 27th International Conference
  on Field Programmable Logic and Applications (FPL)}, 2017, pp. 1--8.

\bibitem{umuroglu2017}
\BIBentryALTinterwordspacing
Y.~Umuroglu \emph{et~al.}, ``{FINN}: A framework for fast, scalable binarized
  neural network inference,'' in \emph{Proceedings of the 2017 ACM/SIGDA
  International Symposium on Field-Programmable Gate Arrays}, ser. FPGA
  '17.\hskip 1em plus 0.5em minus 0.4em\relax New York, NY, USA: Association
  for Computing Machinery, 2017, p. 65–74. [Online]. Available:
  \url{https://doi.org/10.1145/3020078.3021744}
\BIBentrySTDinterwordspacing

\bibitem{zhang2018}
X.~Zhang \emph{et~al.}, ``{DNNBuilder}: an automated tool for building
  high-performance dnn hardware accelerators for {FPGAs},'' in \emph{2018
  IEEE/ACM International Conference on Computer-Aided Design (ICCAD)}, 2018,
  pp. 1--8.

\bibitem{schibisch2018}
S.~Schibisch \emph{et~al.}, ``Online label recovery for deep learning-based
  communication through error correcting codes,'' 2018.

\bibitem{softDemapping}
P.~Robertson \emph{et~al.}, ``A comparison of optimal and sub-optimal {MAP}
  decoding algorithms operating in the log domain,'' in \emph{Proceedings IEEE
  International Conference on Communications ICC '95}, vol.~2, 1995, pp.
  1009--1013 vol.2.

\end{thebibliography}
